\documentclass[sigconf,twocolumn]{acmart}
\usepackage{natbib}

\AtBeginDocument{%
  }

\setcopyright{acmlicensed}
\copyrightyear{2018}
\acmYear{2018}
\acmDOI{XXXXXXX.XXXXXXX}

\acmJournal{JDS}
\acmVolume{37}
\acmNumber{4}
\acmArticle{111}
\acmMonth{8}

\begin{document}

\title{Stories That Teach: Eastern Wisdom for Human-AI Creative Partnerships}

\author{Kexin Nie}
\email{niekexinbella@gmail.com}
\orcid{0009-0002-9190-092X}
\affiliation{%
  \institution{The University of Sydney}
  \city{Sydney}
  \country{Australia}
}

\author{Xin Tang}
\email{xintang0119@gmail.com}
\orcid{0009-0007-7094-4233}
\affiliation{%
  \institution{Guangzhou Academy of Fine Arts}
  \city{Guangzhou}
  \country{China}
}

\author{Mengyao Guo}
\authornote{Mengyao Guo is the corresponding author.}
\email{guomengyao@hit.edu.cn}
\orcid{0009-0009-6016-5900}
\affiliation{%
  \institution{Harbin Institute of Technology, Shenzhen}
  \city{Shenzhen}
  \country{China}
}

\author{Ze Gao}
\email{zegaoap@hotmail.com}
\orcid{0000-0001-9347-6312}
\affiliation{%
  \institution{HongKong Polytechnic University}
  \city{Hong Kong SAR}
  \country{China}
}

\renewcommand{\shortauthors}{Nie et al.}

\acmArticleType{Review}

\acmCodeLink{https://github.com/borisveytsman/acmart}
\acmDataLink{htps://zenodo.org/link}

\acmContributions{BT and GKMT designed the study; LT, VB, and AP
  conducted the experiments, BR, HC, CP and JS analyzed the results,
  JPK developed analytical predictions, all authors participated in
  writing the manuscript.}

\begin{abstract}
This workshop explores innovative human-AI collaboration methodologies in HCI visual storytelling education through our established "gap-and-fill" approach. Drawing on Eastern aesthetic philosophies of intentional emptiness, including Chinese negative-space traditions, Japanese "ma" concepts, and contemporary design minimalism, we demonstrate how educators can teach students to maintain creative agency while strategically leveraging AI assistance. During this workshop, participants will experience a structured three-phase methodology: creating a human-led narrative foundation, identifying strategic gaps, and collaborating on AI enhancements. The workshop combines theoretical foundations with intensive hands-on practice, enabling participants to create compelling HCI visual narratives that demonstrate effective human-AI partnership. Through sequential art techniques, storyboarding exercises, and guided AI integration, attendees learn to communicate complex interactive concepts, accessibility solutions, and user experience flows while preserving narrative coherence and creative vision. Building on our successful workshops at ACM C\&C 2025~\cite{guo2025visual}, this session specifically addresses the needs of the Chinese HCI community for culturally informed and pedagogically sound approaches to AI integration in creative education.
\end{abstract}


\begin{teaserfigure}
  \includegraphics[width=\textwidth]{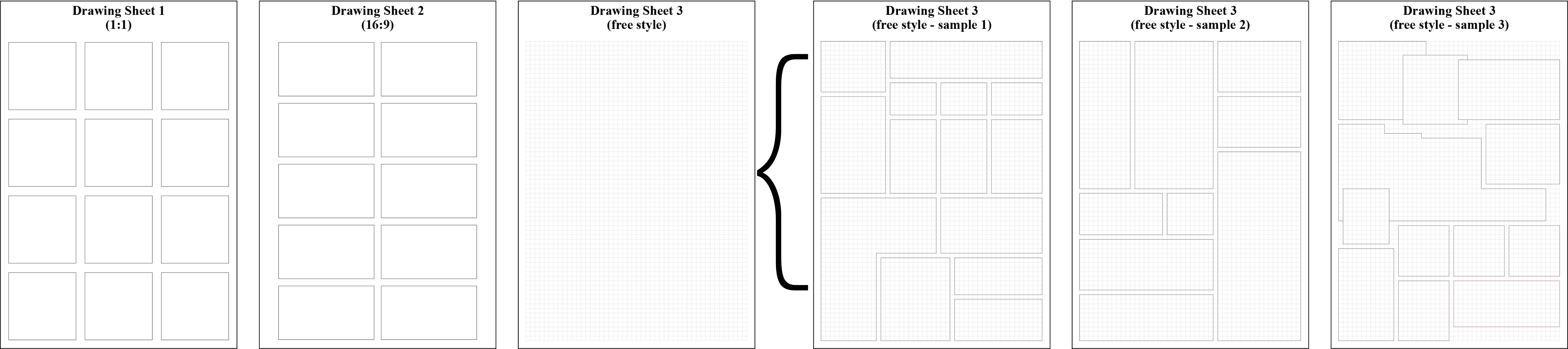}
  \centering
  \caption{The drawing templates for this workshop.}
  \Description{The drawing templates for this workshop.}
  \label{teaser}
\end{teaserfigure}

\maketitle

\section{Motivation \& Objectives}

\subsection{Our Motivation}

The rapid advancement of generative AI has created both opportunities and challenges in the field of HCI education. Chinese universities and research institutions are increasingly integrating AI tools into their curricula, yet many educators struggle to balance human creativity with AI assistance. Students often fall into two extremes: complete dependence on AI generation or complete avoidance of these tools. 

Our motivation stems from: 

\begin{enumerate}
    \item Cultural Relevance: Eastern aesthetic traditions of intentional emptiness provide unique philosophical foundations for human-AI collaboration;
    \item Educational Need: Growing demand for practical frameworks that preserve student creative agency while teaching AI collaboration skills;
    \item Community Gap: Limited resources specifically designed for Chinese HCI educators integrating AI into visual storytelling curricula;
    \item Research Opportunity: Advancing understanding of productive human-AI partnerships in creative educational contexts.
\end{enumerate}

\subsection{Objectives}

By the end of this workshop, participants will: 

\begin{enumerate}
    \item Master the Gap-and-Fill Methodology: Understand and apply our proven framework for human-AI collaborative visual storytelling;
    \item Develop Pedagogical Skills: Create lesson plans and assessment strategies for teaching human-AI collaboration in their own courses;
    \item Build Cultural Connections: Explore how Eastern aesthetic philosophies inform contemporary human-AI creative partnerships;
    \item Create Practical Outputs: Produce HCI visual narratives demonstrating effective human-AI collaboration techniques;
    \item Establish Community Networks: Connect with peers facing similar challenges and opportunities in AI-enhanced creative education;
    \item Advance Research Understanding: Contribute insights to ongoing research on human-AI collaboration in educational contexts.
\end{enumerate}

\section{Organisers \& Expertise}

\textbf{Kexin Nie} is a UI/UX and interactive designer based in Sydney and Beijing, holding a Master's in Strategic Design and Design Innovation from The University of Sydney. Her research expertise centers on human-centered design, Mixed Reality, AI-generated content, and HCI, with particular focus on how AI-driven and immersive interfaces can enhance interaction design methodologies. She brings valuable practical experience in exploring the intersection of emerging technologies and user experience, investigating how contemporary digital tools can be integrated effectively into design education and practice. Her research contributions have been published in prestigious academic venues, including SIGGRAPH Asia, ISMAR, and EMNLP, demonstrating her commitment to advancing the field through rigorous academic inquiry and innovative design solutions.

\textbf{Tang Xin} is a new media artist based in Guangzhou, China, with over 6 years of experience in AI-enhanced Cultural Heritage and Human-Computer Interaction. She is currently pursuing her PhD degree at the Guangzhou Academy of Fine Arts and is known for her contributions to the preservation and promotion of Batik through digital technology. Her works are usually presented in the form of installations and digital media art. Her representative work, AI-Batik, has earned gold awards in the China Brand Design Award and the FA International Frontier Innovation Design Competition, and has been exhibited in Shanghai, Sapporo, and Chonburi. In addition to serving as a board member of ICACHI, she is also involved with the Conference on the Asia Society of Basic Design and Art.

\textbf{Dr. Mengyao Guo} serves as Assistant Professor at Harbin Institute of Technology, Shenzhen, where she specializes in visual communication, digital media arts, and human-computer interaction. She brings exceptional interdisciplinary expertise to visual storytelling methodologies. As an award-winning artist, illustrator, and researcher, Dr. Guo has received numerous international accolades, including the Muse Creative Award, A'Design Award, and DNA Paris Design Award, with her work held in notable collections at K11 Hong Kong and OCAT Shenzhen. Her research has been published in leading venues, including ACM CHI, IEEE VR, ACM IUI, and ACM SIGGRAPH Asia, establishing her as a recognized expert in combining visual arts with cutting-edge technology for innovative HCI applications. She also led the workshop "HCI in Visual Storytelling" successfully at ACM C\&C 2025~\cite{guo2025visual}.

\textbf{Ze Gao} is a multidisciplinary artist and researcher jointly based in Hong Kong and Auckland, serving as Chief Scientist at the Dawa AIGC Research Institute and formerly on the faculty of the Hong Kong Polytechnic University. His scholarship and practice integrate AI-generated content, human–computer interaction, cultural heritage, and mixed reality, yielding publications in premier ACM and IEEE venues (CHI, SIGGRAPH Asia, UbiComp, IEEE VR) and artworks exhibited at SIGGRAPH, ISEA, NeurIPS, CVPR, and DIS, now held by the Rochester Art Center and the Institut Franco-Américain. Recognitions include the Ars Electronica Award, Red Dot Design Award, Lumen Prize, NTU Global Digital Art Prize, and OPPO Renovators Award. He has also served as a reviewer, program committee, and art jury for leading conferences and journals, including ACM CHI, UIST, IEEE VR, ISMAR, PRICAI, ISEA, EVA London, IJHCI, and IJHCS.

\section{Format \& Activities}

\subsection{Phase 1: Foundation \& Context (45 minutes)}

\subsubsection{Opening Presentation (20 min)} 

Organizers introduce the gap-and-fill methodology through visual examples, connecting Eastern aesthetic philosophies (such as Chinese negative space and Japanese "ma") to modern HCI applications. Participants observe how strategic emptiness in visual narratives fosters meaningful AI collaboration opportunities while preserving human creative leadership.

\subsubsection{Cultural Exploration (15 min)} 

Interactive discussion analyzing classical Chinese artworks alongside HCI design examples. Participants explore how intentional gaps invite interpretation and engagement, understanding gap-and-fill as an extension of established aesthetic principles rather than a purely technical approach.

\subsubsection{Framework Introduction (10 min)} 

Clear explanation of the three-phase collaborative approach: human foundation building, strategic gap identification, and AI integration practice. Includes practical examples and success indicators for effective human-AI collaboration.

\subsection{Phase 2: Hands-On Creation (2 hours)}

\subsubsection{Human Foundation Building (45 min)}

Participants create HCI visual narratives using traditional sketching techniques. Focus on narrative structure, character development, and visual storytelling without AI tools. Organizers provide personalized guidance on panel composition and sequential flow.

\subsubsection{Strategic Gap Identification (30 min)} 

Guided exercise teaching participants to recognize appropriate opportunities for AI enhancement. Learn to distinguish between gaps that preserve creative agency versus those that undermine narrative coherence through systematic evaluation techniques.

\subsubsection{AI Integration Practice (45 min)} 

Hands-on experience with AI tools, learning targeted prompting and enhancement techniques. Real-time problem-solving for style consistency, narrative coherence, and effective human-AI collaboration while maintaining creative control.

\subsection{Phase 3: Reflection \& Community Building (45 minutes)}

\subsubsection{Showcase \& Critique (25 min)} 

Participants present their collaborative works (2-3 minutes each), explaining creative processes and integration strategies. Peer feedback focuses on narrative effectiveness and successful collaboration examples.

\subsubsection{Pedagogical Discussion (15 min)} 

Structured sharing of implementation strategies, assessment approaches, and common challenges. Rapid exchange of concrete solutions for curriculum integration and institutional barriers.

\subsubsection{Community Formation (5 min)} 

Establish ongoing collaboration networks through communication channels, contact exchange, and identification of future partnership opportunities for continued support and resource sharing.

\section{Expected Size}

20 participants. This size ensures personalized mentorship during hands-on activities while maintaining group dynamics for meaningful peer discussion and feedback. With 20 participants, organizers can provide individual guidance during the intensive creation phases, facilitate effective small group collaborations, and manage logistics for material distribution and AI tool access.

\section{Audience \& Participation Format}

Our target audience includes HCI educators and university instructors teaching interaction design, user experience, or human-computer interaction who seek practical strategies for integrating AI tools into their curricula while preserving student creative agency. We also welcome computer graphics instructors, creative technology researchers, and art \& design faculty teaching digital art, visual communication, or interactive media. Industry practitioners, such as UX designers and design educators working in technology companies, who have educational interests, are encouraged to participate. Graduate students and early-career researchers exploring human-AI collaboration in creative contexts will find valuable pedagogical insights and practical frameworks for their future teaching endeavors.

Participation is in-person only to ensure effective hands-on collaboration, real-time mentorship, and meaningful community building, which form the foundation of our methodology. Participants should have basic familiarity with visual design principles and prior experience with any AI tools (text-to-image generators, design assistants, etc.), along with either teaching experience in HCI/design fields, research experience in human-AI collaboration, or professional UX/UI practice with educational interests. The workshop requires commitment to post-workshop community engagement and active participation in all three phases, as the collaborative learning model depends on peer interaction and shared problem-solving throughout the session.

\section{Pre-workshop \& course logistics}

\subsection{Required Preparations} 

Participants must bring a laptop or tablet with a reliable internet connection and active accounts for at least one text-to-image generator. Basic drawing software must be installed (Procreate, Adobe Creative Suite, or equivalent free alternatives). Workshop templates and inspiration cards will be emailed one week in advance for download and familiarization.

\subsection{Recommended Preparations}

Two weeks before the workshop, participants will receive reading materials on Eastern aesthetic philosophies and human-AI collaboration, along with a brief reflection questionnaire about current teaching challenges and AI integration goals. A technical support session will be offered 1 week prior to troubleshooting to ensure all software and AI tools function properly on participant devices.

\section{Post-workshop \& course outcomes}

\subsection{Immediate Outputs}

Each participant will create a personal learning portfolio showcasing their visual narratives that demonstrate human-AI collaboration techniques, along with a collaborative collection of pedagogical resources, including lesson plans, assessment rubrics, and implementation strategies developed during group discussions. Workshop documentation will capture community insights, challenges, and solutions for broader dissemination.

\subsection{Medium and Long-term Impact}

An online exhibition will showcase participants' works for one month following the conference, providing lasting visibility for workshop innovations. Follow-up surveys and implementation reports will document how participants integrate the methodology into their teaching, contributing to a comprehensive research publication on workshop findings and pedagogical innovations for submission to relevant journals. A sustained online community platform will enable ongoing resource sharing, troubleshooting support, and collaborative curriculum development, ensuring the workshop's impact extends beyond the immediate session through continued partnerships and follow-up workshops at future conferences based on participant feedback and implementation experiences.

\section{Accessibility considerations}

The workshop ensures inclusive participation through comprehensive accessibility measures, including high-contrast visual materials with alternative text descriptions, live captioning during presentations, and wheelchair-accessible seating with adjustable table heights. For participants with visual impairments, tactile drawing templates, audio descriptions, and screen reader-compatible AI tools will be provided. All digital resources follow WCAG guidelines, assistive technology compatibility is ensured, and specific accommodations can be requested during registration to guarantee full participation regardless of physical, sensory, or cognitive differences.

\section{Call for Participation text}

Join us for an innovative workshop exploring how Eastern aesthetic philosophies can inform effective human-AI collaboration in HCI education. As AI tools become ubiquitous in creative workflows, educators face the challenge of teaching students when and how to collaborate with AI while maintaining creative agency and avoiding over-dependence. Our workshop introduces the "gap-and-fill" methodology, rooted in Chinese negative-space traditions and Japanese "ma" concepts, and offers a culturally informed framework for productive human-AI partnership. Through intensive hands-on activities, participants will create original HCI visual narratives that demonstrate strategic AI integration while preserving human creative leadership.

What You'll Learn: 

\begin{itemize}
    \item Master the gap-and-fill methodology for human-AI collaborative storytelling;
    \item Develop practical pedagogical strategies for AI integration in creative curricula;
    \item Create assessment frameworks that fairly evaluate collaborative work;
    \item Build sustainable community networks for ongoing support and resource sharing.
\end{itemize}

\subsection{Who Should Attend}

HCI educators, interaction design instructors, creative technology researchers, and industry practitioners with educational interests. Basic familiarity with visual design principles and AI tools required.

\subsection{Workshop Format} 

Three intensive phases covering theoretical foundations, hands-on creation, and community building. Participants will work with traditional sketching techniques and AI tools to develop complete visual narratives while learning to identify strategic opportunities for meaningful collaboration.

\subsection{Pre-workshop Requirements} 

Laptop with internet connection, AI tool accounts (Doubao, Midjourney, or similar), and basic drawing software. All physical materials and templates provided.

\bibliographystyle{ACM-Reference-Format} 
\bibliography{main} 

@inproceedings{guo2025visual,
author = {Guo, Mengyao and Nie, Kexin and Han, Jinda and Wang, Guan and Wang, Xin and Chi, zhishun and Fu, Jie and Gao, Ze},
title = {Visual Storytelling in HCI: A Workshop on Narrative Development Through Sequential Art},
year = {2025},
isbn = {9798400712890},
publisher = {Association for Computing Machinery},
address = {New York, NY, USA},
url = {https://doi.org/10.1145/3698061.3728391},
doi = {10.1145/3698061.3728391},
booktitle = {Proceedings of the 2025 Conference on Creativity and Cognition},
pages = {3–12},
numpages = {10},
location = {
},
series = {C\&C '25}
}

\end{document}